\documentclass[runningheads]{llncs}
\usepackage{graphicx}
\usepackage{textcomp}
\usepackage{comment}
\usepackage{todonotes}
\usepackage{color}

\usepackage{array}
\newcolumntype{L}[1]{>{\raggedright\let\newline\\\arraybackslash\hspace{0pt}}m{#1}}
\newcolumntype{C}[1]{>{\centering\let\newline\\\arraybackslash\hspace{0pt}}m{#1}}
\newcolumntype{R}[1]{>{\raggedleft\let\newline\\\arraybackslash\hspace{0pt}}m{#1}}
\begin{document}
\title{Quality-aware semi-supervised learning for CMR segmentation} 

\author{Bram Ruijsink* \inst{1,2,3} \and 
Esther Puyol-Ant\'on* \inst{1} \thanks{Joint first authors.}  \and
Ye Li \inst{1} \and 
Wenja Bai \inst{4} \and 
Eric Kerfoot \inst{1} \and 
Reza Razavi  \inst{1,2} \and
Andrew P. King \inst{1}}
\authorrunning{Bram Ruijsink and E Puyol-Ant\'on et al.}   
\institute{School of Biomedical Engineering \& Imaging Sciences, King\textquotesingle s College London, UK. \and St Thomas\textquotesingle{} Hospital NHS Foundation Trust, London, UK. \and Department of Cardiology, University Medical Centre Utrecht, The Netherlands. \and Biomedical Image Analysis Group, Imperial College London, UK.}

\maketitle              
\begin{abstract} 
One of the challenges in developing deep learning algorithms for medical image segmentation is the scarcity of annotated training data. To overcome this limitation, data augmentation and semi-supervised learning (SSL) methods have been developed. However, these methods have limited effectiveness as they either exploit the existing data set only (data augmentation) or risk negative impact by adding poor training examples (SSL). 
Segmentations are rarely the final product of medical image analysis - they are typically used in downstream tasks to infer higher-order patterns to evaluate diseases. 
Clinicians take into account a wealth of prior knowledge on biophysics and physiology when evaluating image analysis results.
We have used these clinical assessments in previous works to create robust quality-control (QC) classifiers for automated cardiac magnetic resonance (CMR) analysis. In this paper, we propose a novel scheme that uses QC of the downstream task to identify high quality outputs of CMR segmentation networks, that are subsequently utilised for further network training. In essence, this provides quality-aware augmentation of training data in a variant of SSL for segmentation networks (semiQCSeg).
We evaluate our approach in two CMR segmentation tasks (aortic and short axis cardiac volume segmentation) using UK Biobank data and two commonly used network architectures (U-net and a Fully Convolutional Network) and compare against supervised and SSL strategies. We show that semiQCSeg improves training of the segmentation networks. It decreases the need for labelled data, while outperforming the other methods in terms of Dice and clinical metrics. SemiQCSeg can be an efficient approach for training segmentation networks for medical image data when labelled datasets are scarce.\\
\keywordname{ CMR; segmentation network; quality control; data augmentation}
\end{abstract}

\section{Introduction}
Automated segmentation and analysis of medical imaging data using deep learning (DL) has been shown to greatly improve analysis of a wide range of diseases. Using the output of segmentation networks, functional processes in the body can be quantified. For example, in cardiac magnetic resonance (CMR) imaging, the segmentations obtained from individual frames of a cine acquisition of the beating heart can be used to calculate the volumes of the ventricles throughout the cardiac cycle, providing a detailed description of the pumping function of the heart.\\ 
One of the main challenges facing the use of DL models in medical imaging is the scarcity of training data. To createa complete training dataset for cardiac volume segmentation over the full cardiac cycle from short-axis CMR, multiple slices (typically 10-12) and multiple time-frames (typically 30-50) need to be manually segmented for each case by an experienced CMR cardiologist. This labour intensive task cannot be feasibly done at scale. As a result, segmentations are often available in only one or two time-frames.\\ 
Data augmentation 
has become a vital part of network training. Most augmentation techniques focus on the existing training data only, typically applying spatial and intensity transformations. Semi-supervised learning (SSL) techniques have also been proposed \cite{Cheplygina2019}, in which the training data are augmented by new cases labelled by a DL model trained in a supervised way \cite{Bai2017semisup}. However, if the new labels are unreliable this has the potential to degrade model performance.\\
So far, information about downstream tasks has mostly been ignored when producing extra training data for segmentation networks. 
However, such knowledge is potentially very useful: the behaviour of biological segmentation targets is constrained by their biophysical properties. For example, the shape and volume of the aorta and/or cardiac left and right ventricles (LV and RV), and their changes during the cardiac cycle follow certain patterns, even in disease. Clinicians intuitively use this biophysical knowledge when evaluating results of CMR analysis. In previous work, we showed that a clinician's assessment of shape and volume changes during CMR analysis can be used to train quality control (QC) classifiers to identify erroneous segmentations \cite{jacc2018EB}. In this paper, we aim to utilise these QC classifiers to increase the training data of segmentation models in a semi-supervised setting. Specifically, we propose to identify highly accurate segmentation results on unlabelled cases (i.e. `best-in-class' cases) using the QC classifiers and use these cases to increase the available training data.\\
We evaluate our proposed approach in two different segmentation tasks (aortic and cardiac segmentation) using two commonly used architectures (fully convolutional network (FCN) \cite{bai2017semi}
and U-Net \cite{Ronneberger2015}). Our results show that the proposed method outperforms state-of-the-art SSL and fully supervised methods, while significantly reducing the initial (manually labelled) training set needed.\\  
\textbf{Related Work:} Data augmentation techniques can be used to create new and realistic-looking training data. Most augmentation methods involve alteration of existing training examples, for instance by using spatial and intensity transformations or applying kernel filters \cite{2019survey}. More recently, DL approaches have been employed to generate new training data. Several works have proposed synthesis of image segmentation training data using Generative Adversarial Networks (GANs), for example for computed tomography \cite{sandfort2019data} and chest X-rays \cite{bozorgtabar2019informative}.
Another approach, demonstrated by Oksuz \textit{et al.} \cite{oksuz2019automatic}, is to use knowledge of CMR acquisition to synthesise a specific image type of interest. The authors created images that contained motion artefacts by reverse modelling non-corrupted images to k-space and inserting irregularities that would have occurred if motion were present during acquisition.\\ 
SSL methods represent another way of improving training in the face of limited training data. Examples include self-training \cite{Kervadec2019}, bootstrapping \cite{Qiaoet2018} and generative modeling \cite{Hung2018}. Most common approaches focus on weakly supervised learning \cite{Qiaoet2018}, \cite{Radosavovic2018}, which consists of generating pseudo pixel-level labels to increase the training database and then train a segmentation network in a supervised manner. For generative modeling, extensions of the generic GAN framework have been used to obtain pixel-level predictions that can be used as training samples \cite{Hung2018}. In this case the discriminator is jointly trained with an adversarial loss and a supervised loss over the labelled examples.\\
In the cardiac imaging domain, Bai \textit{et al.} \cite{Bai2017semisup}, proposed a SSL approach for CMR image segmentation that combines labelled and unlabelled data during training. This technique used a conditional random field (CRF) based postprocessing step to improve the quality of new training samples produced by a FCN. Another SSL approach was proposed by Kervadec \textit{et al.} \cite{Kervadec2019}, who used a teacher-student network approach to segment 3D atrial volumes. A more general approach to improve segmentation outcomes is by applying multiple networks simultaneously and select only the best result \cite{Hann2019}.\\
The benefits of standard data augmentation are limited by the variation present in the existing data, while DL-generated augmented examples and SSL potentially suffer from generation of unrealistic or inaccurate data. These are significant drawbacks, particularly in medical image analysis. 
Our proposed method utilises aspects of data augmentation and SSL alongside robust QC. It creates an approach for semi-supervised DL model training that is informed by an independent QC network, trained
using prior knowledge of biophysical behaviour in a downstream task. To the best of our knowledge, no previous works have exploited downstream analysis of model output for a QC-informed SSL approach to image segmentation.
\section{Materials}
\label{sec:materials}
We evaluate our proposed approach in two different segmentation tasks using  two separate segmentation models. The first task entails segmentation of the aortic area (1 label) from a cine CMR acquisition of the ascending aorta using a FCN \cite{bai2017semi}.
The second task entails segmentation of the LV and RV blood pools, and LV myocardium (3 labels) from a short-axis (SAX) cine CMR acquisition using a U-net \cite{Ronneberger2015}. The image datasets were obtained from the UK Biobank cohort. The aortic cine CMR acquisitions consists of a single slice and 100 temporal frames. The SAX acquisition contains multiple slices covering the full heart. This acquisition contains 50 temporal frames. Details of the image acquisition protocol can be found in \cite{petersen2016uk}.\\
For training of the models, we used manual segmentations obtained by experienced CMR operators. The training set consisted of 138 subjects with 3 segmented timeframes (randomly chosen throughout the cardiac cycle)
for the aortic images, and 500 subjects with 2 segmented timeframes each for the SAX data (at the end-diastolic and end-systolic phases). This approach is in keeping with common practice for training DL segmentation models for cine CMR. Both datasets consisted of a mix of healthy subjects and patients with a range of cardiovascular diseases.\\
For evaluation, we used a set of test cases with full cardiac cycle ground truth segmentations in a total of 102 subjects for both the aortic images and SAX images. These datasets consist of 50\% healthy subjects and 50\% patients with cardiomyopathies.  
\section{Methods}
\label{sec:methods}
\begin{figure}[ht]
\centering
\includegraphics[width=1.0\textwidth]{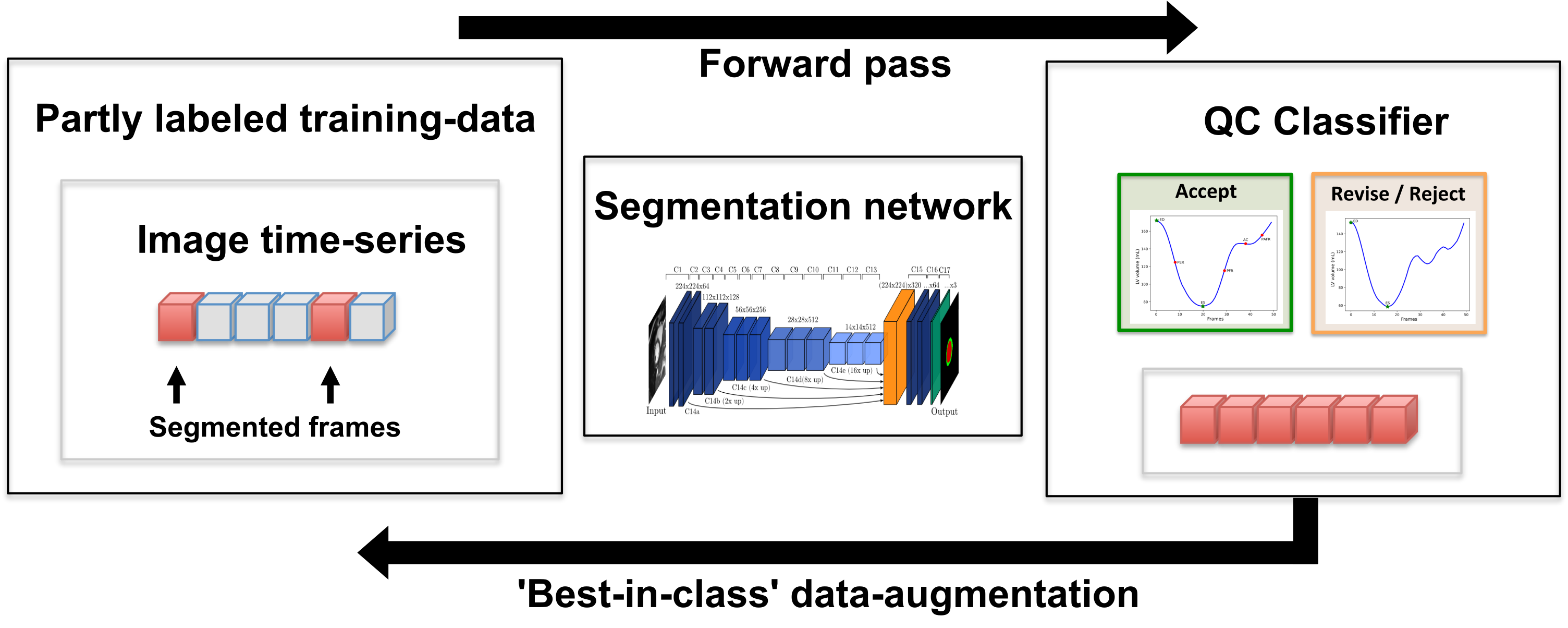}
\caption{Overview of the proposed framework for QC semi-supervised learning.}
\label{fig:pipeline}
\end{figure}

In the following sections we  describe the automated segmentation algorithms, the QC step and the process of identifying the `best-in-class' segmentations. Figure \ref{fig:pipeline} summarises these steps and gives an overview of our proposed method.  

\subsection{Proposed framework}
\label{subsec:clinical_descriptors}
\textbf{Automatic segmentation network:} We used a FCN with a 17 convolutional layer VGG-like architecture for automatic segmentation of the aorta in all frames through the cardiac cycle \cite{bai2017semi,jacc2018EB}. For segmentation of the LV blood pool, RV blood pool and LV myocardium from the cine SAX images, we used a U-net with the layers of the encoding and decoding stages defined using residual units \cite{kerfoot2018left}.\\
\textbf{Output measures:} From each segmentation output, the aortic area and LV and RV blood pool volumes were calculated. By combining the measures from all time-frames, curves of aortic area and ventricular blood volume change were formed.\\  
\textbf{Quality control:} For QC, we used a 3-layer Long-Short-Term-Memory (LSTM) classifier, preceded by a single dense-layer convolution. The aim of this network is to classify cases into `good' and `erroneous' based on the area/volume curves. Separate classifiers were trained for each of the aortic area and ventricular volume data. To train the networks, a clinician reviewed 1,000 cases using plots of the LV and RV blood volume curves and aortic area change in combination with an animation of the obtained segmentations, see also \cite{jacc2018EB}.
The 1,000 training cases for QC were randomly selected from the UK Biobank dataset and were not used for training or evaluating our segmentation models. This way, overfitting of the QC network to the segmentation training data was avoided.\\
The loss functions of these networks were optimised for sensitivity of detection of erroneous results. This was done by obtaining a receiver operating characteristic (ROC) curve after training and selecting the classifier that maximised sensitivity (i.e. recall) using a weighted Youden index. The QC classifiers used during the experiments described in this paper had a sensitivity of 96\% (aorta) and 95\% (SAX) for detection of erroneous results.\\
\textbf{Identifying `best-in-class' cases:} After QC, we used the probability outputs of the QC classifiers to rank cases from highest probability of being accurate to lowest probability of being accurate. We identified the 30 `best-in-class' cases and from these cases, the full-cardiac cycle segmentations and their corresponding images were used to augment the training dataset of the segmentation model.
\subsection{Experiments}
\textbf{Network training:} We trained both segmentation models in four scenarios: (1) using all cases originally available for training of the network (i.e. the `full dataset') - this acts as a `best achievable' scenario, (2) a baseline approach using only half the available training data in a fully supervised manner (`half dataset'), (3) using a SSL approach, initially using half of the training data, but feeding the outputs of 30 random cases back into the training set without QC (semi-supervised no QC), (4) a state-of-the-art SSL approach \cite{Bai2017semisup} that utilises CRF postprocessing to improve the quality of the new training segmentations, and finally (5) using our proposed approach (semiQCSeg), initially using half of the training data, and then after feeding the 30 `best-in-class' outputs back into the training data.
Table \ref{table:data} shows a summary of the data used for each experiment. To achieve  a fair comparison, we trained all networks in a similar way: from scratch and for a fixed number of 2,000 epochs.

\begin{table} [!h]
\centering
\caption{Summary of the training data (number of subjects and number of labelled images) used for each experiment. Note that, for the SSL approaches (3)-(5), the number of images increases dramatically because segmentations are available for all timeframes.}
\begin{tabular}{lcccc}
\hline
&  \multicolumn{2}{c}{\textbf{Aorta FCN}} &  \multicolumn{2}{c}{\textbf{SAX U-net}} \\
Methods & subjects (\it{n}) & images (\it{n})  & subjects (\it{n}) & images (\it{n})    \\ \hline 
(1) Full dataset & 138 & 414 & 500 & 1000 \\
(2) Half dataset & 69 & 207 & 250 & 500 \\
(3) Semi-supervised no QC & 99 & 2160 & 280 & 2000 \\
(4) Semi-supervised no QC + CRF& 99 & 2160 & 280 & 2000 \\
(5) SemiQCSeg & 99 & 2160 & 280 & 2000 \\ \hline
\end{tabular}
\label{table:data}
\end{table}

\noindent\textbf{Evaluation:} We evaluated the models' segmentation performances using the independent test set (described above), which was not used during training. We compared segmentation performance  using mean Dice scores for aorta and SAX segmentations and additionally evaluated the performance of the different approaches in the downstream task for the SAX dataset, i.e. the calculation of ventricular volume and ejection fraction.

\section{Results}
\begin{figure}[ht]
\centering
\includegraphics[width=1.0\textwidth]{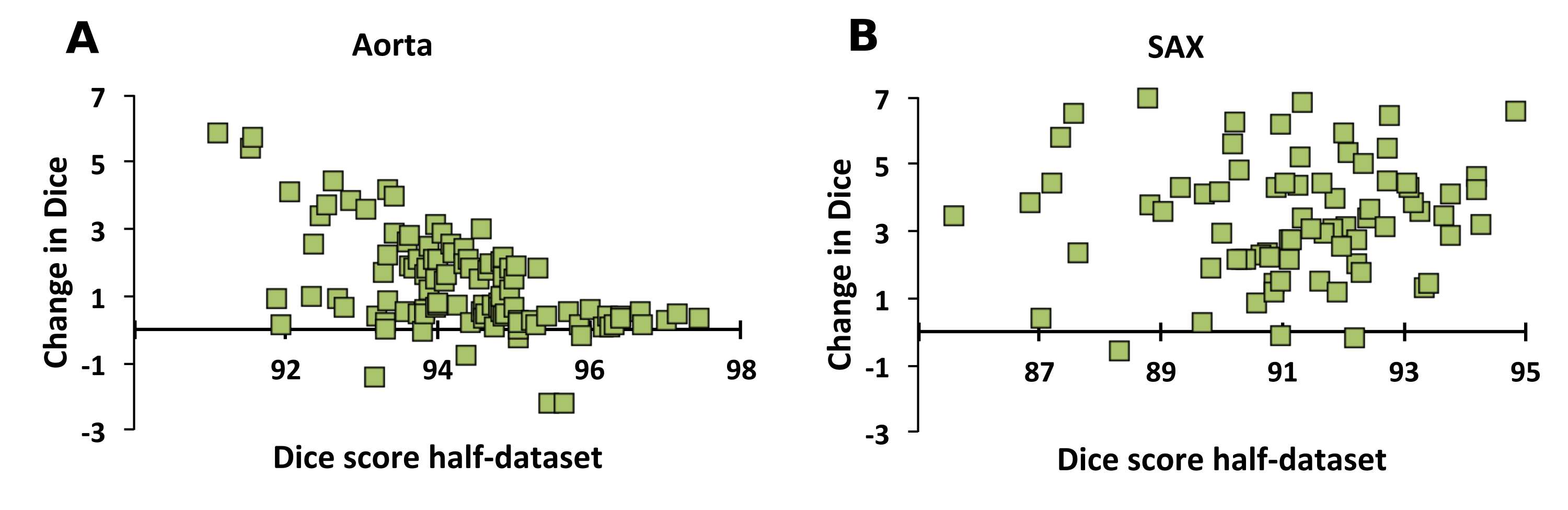}
\caption{Change in Dice score using semiQCSeg with respect to half dataset training for each individual case. A) for aorta, B) for SAX (mean pooled dice of LV and RV blood pool, and myocardium).}
\label{fig:dicedelta}
\end{figure}

\begin{figure}[ht]
\centering
\includegraphics[width=0.9\textwidth]{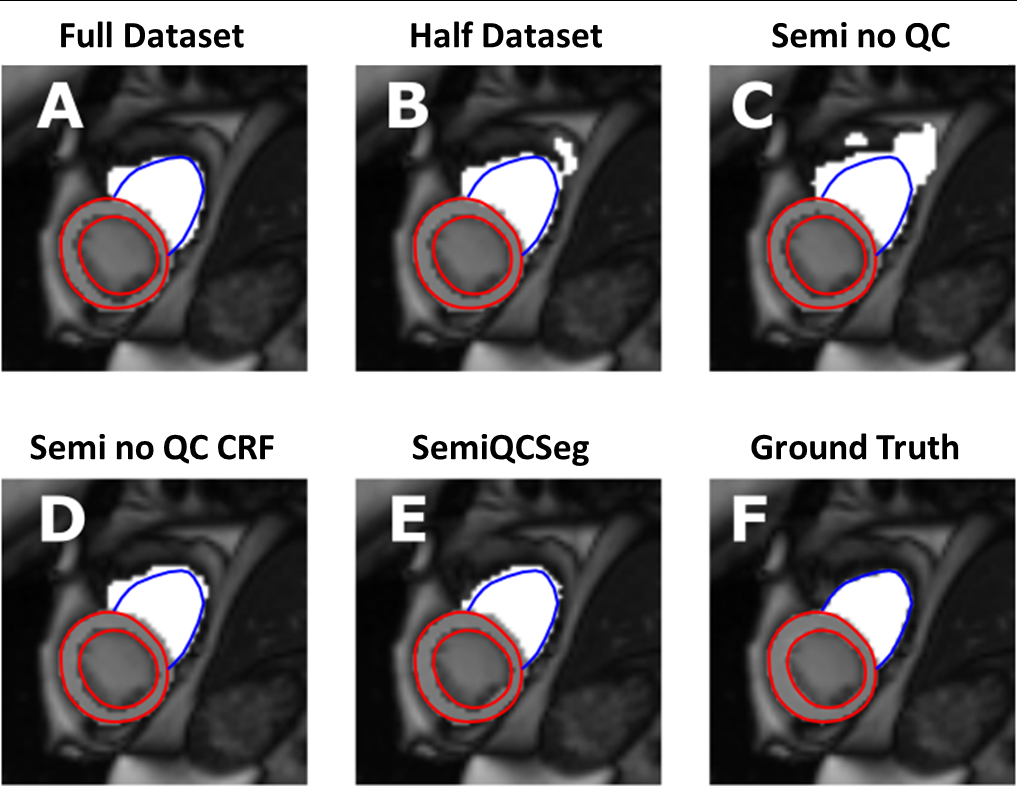}
\caption{Examples of segmentations obtained using the different approaches for a CMR short axis image, with our proposed method (SemiQCSeg) in panel E. The contour overlay in each image represents the ground truth segmentation. QC: quality control, CRF: conditional random field.}
\label{fig:sax_examples}
\end{figure}

Table \ref{table:dice} shows the mean Dice scores obtained using the different approaches. The Dice score shown for the SAX data is a pooled mean of the individual LV and RV blood pool and LV myocardium values.\\
The semiQCSeg approach achieved a significantly higher Dice score than the `half dataset' baseline approaches for both SAX and aorta experiments. More importantly, semiQCSeg outperformed the best achievable model (`full dataset') for SAX, while achieving Dice scores similar to full dataset training for Aorta. SSL without QC performed significantly worse than our semiQCSeg approach. Adding the CRF postprocessing to this model improved performance, but not to the level of our proposed approach. See Figure \ref{fig:sax_examples} for an example of the performance of the different approaches for segmentation of a SAX stack of a selected subject.\\ 
Selecting output cases for training augmentation might bias a network towards learning to segment ‘easy’ cases only, while decreasing accuracy in ‘hard’ cases. Figure \ref{fig:dicedelta} shows that there was no bias in improvement of Dice scores in easy cases alone using semiQCSeg.

\begin{table} [!h]
\centering
\caption{Mean Dice and standard deviation (between brackets) of the tested models. For SAX, the pooled mean dice of LV and RV blood pool and LV myocardium is shown. \textsuperscript{*} denotes a \textit{p}-value$<$.001 with respect to semi-supervised QC approach (semiQCSeg), using a paired t-test.}
\begin{tabular}{L{5cm}C{2.5cm}C{2.5cm}}
\hline
& \textbf{Aorta FCN} & \textbf{SAX U-net} \\
Training strategy & Dice & Dice  \\ \hline 
(a) Full database & 95.45 (1.45) & 88.17 (4.35)\textsuperscript{*}  \\
(b) Half database & 93.71 (1.51)\textsuperscript{*} & 82.57 (10.79)\textsuperscript{*} \\
(c) Semi-supervised no QC & 92.86 (2.95)\textsuperscript{*} & 85.30 (6.97)\textsuperscript{*} \\
(d) Semi-supervised no QC CRF & 94.89 (1.71)\textsuperscript{*} & 85.97 (6.00)\textsuperscript{*} \\
(e) SemiQCSeg & 95.56 (1.53)& 89.44 (4.01) \\
\end{tabular}
\label{table:dice}
\end{table}

\subsection{Downstream task}
 A comparison of the different approaches in terms of downstream clinical metrics (ventricular volumes and ejection fraction) for the SAX dataset is shown in Table \ref{table:clinical}. Of all the methods, semiQCSeg yields results closest to those based on manual segmentations. We do not show data for the aortic cases, as the calculation of the clinical metric (aortic distensibility) necessitates additional blood pressure data.


\section{Discussion}
\label{sec:discussion}
In this paper, we have proposed a novel method for quality-aware SSL for medical image segmentation.
We show that this approach is able to boost network performance, while markedly reducing the amount of annotated datasets needed.\\ 
\begin{table} [!h]
\centering
\scriptsize
\caption{Mean and standard deviation (in  brackets) of clinical metrics produced by the tested models. LVEDV: Left Ventricular End Diastolic Volume, LVEF: Left Ventricular Ejection Fraction, RVEDV: Right Ventricular End Diastolic Volume, RVEF: Right Ventricular Ejection Fraction.}
\begin{tabular}{L{4cm}C{2cm}C{2cm}C{2cm}C{2cm}C{2cm}C{2cm}}
\hline
\textbf{ } &   \multicolumn{3}{c}{\textbf{SAX U-net}} \\
Training strategy & LVEDV & LVEF & RVEDV & RVEF \\ \hline 
(a) Full database &  151 (36) & 56 (8) & 163 (42) & 52 (6) \\
(b) Half database & 153 (38) & 54 (6) & 171 (46) & 47 (8) \\
(c) Semi-supervised no QC & 152 (39) &  55 (6) & 165 (44) & 47 (8)\\
(d) Semi-supervised no QC CRF & 151 (38) & 56 (6) & 164 (42) & 48 (7) \\
(e) SemiQCSeg & 150 (36) & 57 (6) & 160 (43) & 54 (6)\\
(f) Manual segmentation & 147 (36) & 58 (5) & 156 (42) & 55 (6)\\
\end{tabular}
\label{table:clinical}
\end{table}
Importantly, our approach does not infer quality from the individual segmentations directly (for example using voxel probability). Instead, it utilises clinical characteristics obtained from a downstream analysis of the segmentations to make use of biophysical knowledge used by clinicians when judging CMR segmentation output. We show that our approach allows segmentation models to learn more effectively; it uses less initial data, while yielding similar or even better results compared to full dataset training. The latter illustrates the strength of our approach; it delivers extra full cardiac cycle datasets while tightly controlling their quality using metrics not directly obtained from the segmentation network.\\
Augmenting data based on direct inference from the segmentations may bias further learning towards a certain group of cases. As can be appreciated in Figure \ref{fig:dicedelta}, this bias seems not to be present using our proposed approach.  However, more robust evaluation is needed to ensure the impact of our approach on bias after multiple iterations of the semiQCSeg process. A large number of iterations is likely to result in some bias in learning. Therefore this method is most suitable for boosting performance using a single or limited number of iterations. To detect increasing bias over iterations, we propose to monitor changes in loss of the validation set for a potential increase to trigger early stopping of the semiQCSeg process.  \\
While training of our QC step necessitates some extra effort from clinicians, this pattern recognition task is fast compared to the time needed for extra segmentation. Moreover, as it is independent of the upstream segmentation network, the trained QC model could be reused flexibly for training different CMR datasets or other segmentation models.\\
We found that SSL without QC achieved some improvement in Dice scores with respect to half dataset training for SAX, but not for the aorta dataset. Including CRF postprocessing led to improvements in both datasets (Table \ref{table:dice}). This is in keeping with earlier reports of SSL approaches \cite{Cheplygina2019}. However, the variability of the impact in aorta and SAX illustrates the uncertainty in uncontrolled data augmentation in SSL; the benefit or negative effect depends entirely on the quality of the output segmentations that are obtained for further training.\\
Figure \ref{fig:sax_examples} shows an example of segmentations obtained using the different approaches. This further illustrates the uncertain impact of SSL without QC; as the segmentations obtained (panel C) are worse than the ones obtained using half dataset training alone (panel B). Of all methods, SemiQCSeg (panel E) performed the best compared to the ground truth (panel F).\\
In this paper, we demonstrated our method in time-series CMR data. The success of our approach opens up the potential to use time-sensitive architectures, such as Long Short Term Memory models, without the need to provide fully labelled training examples that include each timeframe. Extension of our approach to this domain will be the subject of our future work.\\
Our approach could also be used in other segmentation tasks. For example, in computed tomography (multi-)organ segmentation, where segmenting only part of all available slices could be combined with semiQCSeg to provide whole body training sets, or brain imaging, where information obtained from multiple acquisitions can be exploited during QC.\\
In conclusion, we have shown that semiQCSeg is an efficient method for training DL segmentation networks for medical imaging tasks when labelled datasets are scarce.

\section*{Acknowledgements}
Dr. Bram Ruijsink is supported by the NIHR Cardiovascular MedTech Co-operative awarded to the Guy’s and St Thomas’ NHS Foundation Trust. This work was further supported by the EPSRC (EP/R005516/1 and EP/P001009/1), the Wellcome EPSRC Centre for Medical Engineering at King’s College London (WT 203148/Z/16/Z) and National Institute for Health Research (NIHR). This research has been conducted using the UK Biobank Resource under Application Number 17806.\\

\bibliographystyle{splncs04}
\bibliography{refs}

\end{document}